\documentclass [12pt]{article}
\usepackage{axodraw}
\usepackage[pctex32]{graphics}
 \usepackage{graphicx}
 \usepackage{amsmath, amssymb}
\usepackage{array,dcolumn}
\begin{document}
\begin{center}
\small
 { \bf  PRODUCTION AND DECAY OF SAXION IN $e^+e^-$ COLLISIONS }\\[7mm]
\normalsize {\bf Le Nhu Thuc}$^{(a)}$ and
{\bf Dang Van Soa}$^{(a,b)}$\\[1mm]
\small
\textit{$^{(a)}$ Hanoi University of Education, 136 Xuan Thuy, Cau Giay, Hanoi.\\[1 mm]
$^{(b)}$Abdus Salam International Centre for Theoretical Physics, Trieste, Italy}\\[1 mm]
\end{center}
\normalsize
\vspace{9mm}
\begin{abstract}
 The production of saxion in $e^{+}$$e^{-}$ collider
 and the saxion decay  into two photons were calculated in detail. Based on results it shows that
the saxion is stable in our universe and can play the role of the
late decaying particle  ( LDP) in a dark matter.
\end{abstract}
\vspace{1cm}
\section{Introduction}
$\hspace {0.5cm}$
%The Strong-CP problem was proved by Peccei and Quinn (PQ)
The most attractive candidate for the solusion of the strong CP
problem is Peccei and Quinn (PQ) mechanism~\cite{pec}, where the
CP-violating phase $\theta$ ($\theta$$\leq$ $10^{-9}$ ) is
explained by  the existence of a new pseudo-scalar field, called
the axion. Based on the recent laboratory researches and
astrophysical and cosmological considerations~\cite{tur} the value
of the axion mass was estimated in range between $10^{-6}$ eV and
$10^{-3}$ eV~\cite{raf}. The axion appears in different models. In
particular, it appears as a new phase of Higgs fields in the
electroweak theories, or appears as a
term of chiral superfields in the low - energy supersymmetry (SUSY) theories~\cite{raf,kim}.\\
\hspace{0.5cm} The nature of the dark matter in the Universe
 remains one of the most challenging problems in cosmology.
 Numerous candidates for DM have been proposed in the literature.
 One of the most popular candidates in the context of
 supersymmetric theories with R - parity conservation is the
 lightest supersymmetric particle (LPS), namely
 the lightest neutralino. The interactions of the neutralino are weak
 , and its number density at decoupling is therefore often of
the required order of magnitude, which makes it an excellent
candidate
 for the weakly interacting massive particle ( WIMP)~\cite{lau}.

$\hspace*{0.5cm}$ In the SUSY extension of the axion model, the
axion supermultiplet ($\Phi$ = 1/$\sqrt2$(s +ia + $\sqrt 2$$\tilde
a$$\theta$ +
 $F_{\Phi}$$\theta$$\theta$)
consists of the axion, its real scalar superpartner saxion (s),
 and the fermionic superpartner axino ($\tilde a$ ).
 Like axions, the coupling of
the axino to ordinary matter is very weak~\cite{sik}, thus it is a
good candidate for WIMP. The stable relic axino shows that the
axino can be an attractive candidate for cold dark matter
(CDM)~\cite{cov}. Important properties of the axino have bee
studied~\cite{chun1, kim2,dan}. In SUSY axion models, The saxion
mass depends on the specific forms of the axion sector
superpotential, which is predicted in range between $1$ KeV and
$100$ MeV~\cite{chang}. The decay of saxion into two axions with
high values of the Hubble constant is presented in~\cite{asa}. The
saxion properties and its contribution on the dark matter were
studied recently~\cite{got,lyt,chun}. The direct detection of
supersymmetric DM via scattering on nuclei in deep-underground,
low-background experiments has been discussed
many times~\cite{eli,eli1,eli2}.\\
\hspace*{0.5cm}The possible consequences of the presence of
saxions are the subject of this study.
 In this paper we evaluate the production of saxion in the $e^{+}$$e^{-}$
 collision and the saxion decay into two photons.
  The results showed that  the saxion
 is stable in our unverse and it
can play the role of LDP in the dark matter. This paper is
organized as follows: In Sec.II we give constraints on the saxion
mass in the SUSY axion model. In Sec. III and Sec. IV, we evaluate
the production and decay of the saxion in the $e^{+}$$e^{-}$
collisions. Finally, our conclusions are summarized in the last
section.
\section{ Constraint on the saxion mass}
\hspace*{0.5cm}In the framework of gauge mediated SUSY breaking
theories, a gauge singlet PQ multiplet X and colored PQ quark
multiplets $Q_{P}$ and $\overline{Q}_{P}$ were introduced with the
 superpotential~\cite{asa}
\begin{equation}
{W = {\lambda_{P}} X {Q_{P}\overline{Q}_{P}}}
\end{equation}
where $\lambda_{P}$ is a coupling constant here mass parameter was
not introduced. Because in a supersymmetric limit the $U(1)_{PQ}$
symmetry is enhanced to its complex extension, there appears a
flat direction $Q_{P}$ = $\overline{Q}_{P}$ = 0 with X undermined
in the same limit. The balance of the SUSY breaking effects
between the gravity mediation and the gauge mediation stabilizes X
and gives
 the nonzero vacuum expectation value (VEV)  $<$ X $>$ = $F_{a}$ approximately as
\begin{equation}
F_{a} = \frac{f^2}{m_s}
\end{equation}
where $m_{s}$ denotes the mass of a saxion field, the real part of
the scalar component of X, and the mass of saxion is estimated as
$m_{s}$ = $\xi$$m_{3/2}$ with a parameter $\xi$ of order unity.
And $f$ is a mass scale, with a current scalar lepton mass limit,
suggests that $f$ $\gtrsim$$10^{4}$ GeV. The axion arises when the
X field develops the non-vanishing VEV. The decay constant of the
axion (the PQ scale) $F_{a}$ is constrained by various
astrophysical and cosmological consideration, the allowed region
for the PQ scale is discussed~\cite{kim2,asa}
\begin{equation}
10^{9} GeV \lesssim F_{a} \lesssim 10^{12}(10^{16}) GeV
\end{equation}
From Eqs.(2) and (3) we can deduce the allowed region for the
saxion mass, with the minimum value $f$ = $10^{4}$GeV as
\begin{equation}
100 MeV \gtrsim m_{s} \simeq m_{3/2} \gtrsim 1 keV
\end{equation}
Thus the model gives a very simple description of the PQ breaking
mechanism  solely governed by the physics of the SUSY breaking. In
this model it makes sense to consider the possibility of the very
small saxino mass, which is comparable to the gravitino
mass.\\

\section{The production of saxion in $e^{+}$$e^{-}$ collisions}

\hspace*{0.5cm}For the  saxino - photon system, a suitable
Lagrangian density is given by~\cite{lyt,chun}
\begin{equation}
{\cal L}_{(s,\gamma,\gamma)} = \frac{\alpha_c}{8\pi F_a}s F_{\mu\nu}F^{\mu\nu}
\end{equation}
where $\alpha_c$ is the colour constant, $F_{\mu\nu}$ =
$\partial_\mu$$ A_\nu $ - $\partial_\nu$$ A_\mu $
is the field strength tensor.\\
From (5) we get a saxion - photon- photon vertex
\begin{equation}
V_\alpha^{\beta}(s,\gamma,\gamma) = \frac{-i\alpha_c}{4\pi F_a}[
2(q.k)g ^{\beta}_{\alpha} -q^{\beta}k_\alpha - q_\alpha k^{\beta}]
\end{equation}

\hspace*{1cm} Considering the collider process in which the
initial state contains the electron and the positron and the final
state, are the photon and the saxino: {$e^{-}$($k_1$) + $e^{ +
}$($k_2$) $\rightarrow$ $\gamma$($q_1$) + s($q_2$)},
 four-momenta of particles, respectively. This process proceeds
 through the s - channel photon exchange. Investigation on the center-of-mass frame,
 and denoting the scattering angle by $\theta$ ( the angle between momenta of
the initial electron and the final
 photon), where $s = p^2 = (q_1+q_2)^2 = (k_1+k_2)^2$ is
 the square of the collision energy. \\
\hspace*{0.5cm}Supposing that the production of the photon -saxino
pairs at high energies i.e., $m_e, m_s\ll \sqrt{s}$, then the
amplitude for this process is given by
\begin{equation}
\ <f\left | M \right| i\ >= \frac{-ie\alpha_c}{4\pi
F_a}\frac{1}{p^2}  \overline{v}(k_{2})\gamma^{\nu}u(k_{1}) [
2(p.q_1)g_{\nu}^{ \alpha}- p^{\alpha}{q_1}_\nu - p_\nu
q_1^{\alpha}]\epsilon_{\alpha}(q_1)
\end{equation}
The straightforward calculations yields the following differential
cross-section (DCS) as
\begin{equation}
\frac{d\sigma(e^{+}e^{-}\to\gamma s)}{d\Omega} = 4.9 \times
10^{-4}\frac{\alpha \alpha^{2}_{c}} {\pi^{3}F^{2}_{a}} (7 +
cos^{2}\theta),
\end{equation}
where $\alpha$ = $e^{2}$/${4\pi}$ is the structure constant.\\
\begin{figure}[h]
\centerline {\epsfxsize=8cm\epsffile{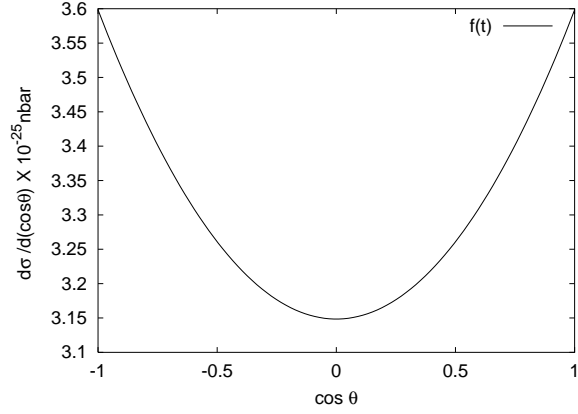}}
 \caption
{\label{Figure.1}{\em DCS }for $e^+e^-\to{\gamma}s$ as a function
of cos$\theta$}
\end{figure}
After integration over the $\theta$ angle, we obtain the total
cross-section as
\begin{equation}
\sigma(e^{+}e^{-}\to\gamma s) = 1.4 \times 10^{-2} \frac{\alpha
\alpha^{2}_{c}}{\pi^{2} F^{2}_{a}}
\end{equation}
 \hspace{0.5cm} From (9) it shows that at high energies cross section ($\sigma$) only
 depends quadratically on PQ scale $F_a$ and $\alpha_c$. With $\alpha_c$ = 0.1, $F_a$ = $10^{11}GeV$
 ~\cite{lyt,chun} and
$\alpha^{-1}$ = 137.0359895, $[GeV]^{-2}$ = 0.39$\times$$10^{-27}$
$cm^{2}$~\cite{tl9}, the total cross-section $\sigma$
($e^{+}$$e^{-}$$\rightarrow$ $\gamma$$s$)
 = 8.3 $\times$ $10^{-24}$ nb. In Fig.1  DCS was plotted by
 $\cos\theta$, as we can see from the figure, DCS is peaked in
 backward and forward direction,
 $\frac{d\sigma(e^{+}e^{-}\rightarrow\gamma s)}{d\Omega} = 3.6\times
 10^{-25}$ nb. The axino production in $e^+e^-$ and $\gamma\gamma$
 collisions is presented in~\cite{soa}.
From our results, it shows that  cross - sections for the saxion
and axino  production at high energies are
 very small, much below neutrino production cross sections, so that the direct production of
 CDM particles is in general not expected
 to lead to easily observable signals in   $e^+e^-$
 annihilation. Note that the experimental upper limit on the DM scattering cross section
 recently provided by the CDMS II as $\sigma\sim 10^{-6}$ pb~\cite{ake}.

\section{The decay of saxion into two photons}

\hspace*{1cm}If the saxino is light enough in which it's mass is
comparable to gravitino mass, then the saxino can not decay into
two gluons but decays into two photons.
 In this section we calculate the decay of saxion into
two photons. The amplitude for this process is
\begin{equation}
\ <f\left | M \right| i\ >= \frac{i\alpha_c}{4\pi F_a}[
2(q.k)g_{\alpha \beta} - k_{\alpha} q_{\beta}- q_{\alpha}
k_{\beta}]\epsilon^{\alpha}(q)\epsilon^\beta(k).
\end{equation}
After some calculations, we obtain the decay rate
\begin{equation}
\Gamma(s\to\gamma\gamma) = 4.9\times 10^{-3} \frac{ \alpha_c^{2}}{
\pi^{3} F^{2}_a}m_s^{3}
\end{equation}
 With $\alpha_c$ = 0.1, $F_a$ = $10^{11}GeV$ and note that
$[GeV]^{1}$ = $\frac{1}{6.6\times 10^{-25}}$$ sec^{-1}$
~\cite{tl9}, then we can rewrite (11) as
\begin{equation}
\Gamma(s\to\gamma\gamma) =
2.3\times10^{-4}\biggl(\frac{\alpha_c}{0.1}\biggl)^{2}
\biggl(\frac{10^{11}GeV}{F_a}\biggl)^{2}m_s^{3}(GeV) sec^{-1}
\end{equation}
Therfore the lifetime of saxion is
\begin{equation}
\tau(s\to\gamma\gamma) = \frac{1}{\Gamma(s\to\gamma\gamma)} =
4.3\times 10^{6}sec^{1}\biggl(\frac{\alpha_c}{0.1}\biggl)^{-2}
\biggl(\frac{F_a}{10^{11}GeV}\biggl)^{2}\biggl(\frac{100MeV}{m_s}\biggl)^{3}
\end{equation}
\hspace*{1cm}For the saxion mass in region
$100MeV$$\leq$$m_s$$\leq$$1 KeV$, the dependence of the decay rate
and the lifetime of the saxion on
its mass is shown in Table 1\\[1.5cm]
%\vspace*{0.1cm}
\begin{center}
 \begin{tabular}{|c|c|c|c|c|c|c|}
\hline
  $m_s(MeV)$&100 & 10  & 1 & 0.1& 0.01 & 0.001\\
\hline
 $\Gamma (sec^{-1})$ &$2.3\times10^{-7}$ & $2.4 \times 10^{-10}$ & $2.3 \times 10^{-13}$ & $2.3\times
 10^{-16}$& $2.3\times 10^{-19}$ & $2.3 \times 10^{-22}$\\
 % &$2.4\times10^{-7}$ \\
\hline $\tau(sec)$ & $4.3 \times 10^{6}$ & $4.3\times 10^{9}$ &
$4.3\times 10^{12}$ &$4.3\times 10^{15}$ & $4.3 \times 10^{18}$ &
$4.3 \times 10^{21}$ \\
%& $4.2 \times 10^{6}$\\
\hline
\end{tabular}
%\vspace{0.3 cm}
\end{center}
\begin{center}
{\bf Table 1}
\end{center}
\hspace*{0.5cm} From the Table 1, we can see that the saxion can
play the role of a long lived constituent of dark matter. When
$m_s\leq 10$ MeV, then  the lifetime of saxion is very long,
therefore the saxion can be a good candidate for LDP. If the
 saxion mass is smaller than 10 KeV , it becomes stable within
 the age of universe ($\sim 10^{17}$ $sec$), when the saxion lifetime is larger the age of the
 universe then saxion oscillation still exists now.
 Note that the value $\tau$ varies very widely because
of the strong dependence on the saxion mass. \\
\hspace*{0.5cm} Now we estimate the cosmic temperature at the
saxion decay ($T_D$)~\cite{asa}\\
\begin{equation}
T_D = 0.6\times A\biggl(\frac{M_G}{A^2 \tau_\sigma}\biggl)^{2/3},
\end{equation}
where A is the saxion abundance: $A\leq 3.6 \times 10^{-9} h^2$,
$M_G = 2,4.10^{18}$ is the Planck scale, $h = 0.7$ is the Hubble
parameter. For $ m_{s}= 10$ KeV  we have $ T_D \sim 726,94$ GeV.

\section{Conclusion}
\hspace*{0.5cm}In our work, we evaluated the production and decay
of saxion in $e^{+}$$e^{-}$ collisions. From our results, it shows
that cross - sections for the saxion  production at high energies
are very small, so that the direct production of
 saxions is  not expected
 to lead to easily observable signals in   $e^+e^-$
 annihilation.\\
\hspace*{0.5cm}The decay of saxion into two photons was also
calculated in detail.
 Based on the results  it shows that the saxion
can play a role of the late decaying massive particle.
 We point out that in all possible
modes the decay of saxion strongly depends on it's mass. If the
mass of saxion is smaller than 10 keV ( correspond to the low
bound of PQ scale ), then the saxion becomes stable and can be a
new candidate for CDM of our universe.\\[0.5cm]
{\it Acknowledgments}\\
\hspace{0.5cm}D. V. Soa would like to thank the Abdus Salam
International Centre for Theoretical Physics (ICTP) for
hospitality and support. This work was supported in part by the
Vietnam National Basic Research Programme on National Science
under grant number KT 410604.\\[1.5cm]

%{\bf Appendix}\\
%\hspace{0.5cm}The saxion has a coupling to the axion through
%Lagrangian\\
%\begin{equation}
%L_{eff} = \frac{f}{F_a}s
%\end{equation}

\begin{thebibliography}{9}
\bibitem{pec}  R. D Peccei and H. R.Quinn, Phys. Rev. D {\bf 16}, 1791
(1977).
\bibitem{tur}  M. S. Turner, Phys. Rep. {\bf 197}, 67 (1990).
\bibitem{raf}  G. Raffelt, Phys. Rep. {\bf 198}, (1990).
\bibitem{kim} J. E. Kim, A. Masiero and D. V. Nanopoulos, Phys. Lett. B{\bf 139}, 346
(1984).
\bibitem{lau} Laura Covi, L. Roszkowski, and M. Small, JETP
{\bf 0207}, 023 (2002).
\bibitem{sik} P. Sikivie, Nucl. Phys. Proc. Suppl. {\bf 87}, 41 (2000).
\bibitem{cov} L. Covi, H. B. Kim, and L. Roszkowski, {\it J. High
Energy Phys}, {\bf 0105}, 033 (2001).
\bibitem{chun1} E. J. Chun, H. B. Kim, and D. H. Lyth, Phys. Rev. D{\bf 62}, 125001 (2000).
\bibitem{kim2} H. B. Kim and J. E. Kim, Phys. Lett. B{\bf 527}, 18 (2002).
\bibitem{dan} D. Hooper and L. T. Wang, Phys. Rev. D {\bf 70}, 063506 (2004).
\bibitem{chang} S. Chang and  H. B. Kim, Phys. Rev. Lett. {\bf 77}, 591(1996).
\bibitem{asa}  T. Asaka and M. Yamaguchi. Phys. Rev. D{\bf 59},125003(1999).
\bibitem{got} T. Goto and M. Yamaguchi, Phys. Lett. B{\bf 276}, 103 (1992);
 E. J. Chun, J. E. Kim and H. P. Nilles,
Phys. Lett. B{\bf 287},
 123 (1992).
\bibitem{lyt} J.E. Kim, Phys. Rev. Lett. {\bf 67}, 3465 (1991); D. H. Lyth, Phys. Rev. D{\bf 48}, 4523 (1993).
\bibitem{chun}  E. J. Chun and A. Lukas, Phys. Lett. B{\bf 357}, 43(1995).
\bibitem{eli} J. Ellis, A. Ferstl, and K. A. Olive, Phys. Lett. B{\bf
532}, 318(2002).
\bibitem{eli1} J. R. Ellis, K. A. Olive, and Y. Santoso, Phys. Rev. D{\bf
67}, 123502(2003).
\bibitem{eli2} J. R. Ellis, K. A. Olive, Y. Santoso, and Vassilis C. Spanos, Phys. Rev. D{\bf
71}, 095007(2005).
\bibitem{soa} H. N. Long and D. V. Soa, Preprint, ICTP,
IC/2000/182. Comm. in Phys., Vol. 13, No.4, 245(2002).
\bibitem{tl9}  H. Q. Kim and P. X. Yem {\it" Elementary Particles and Their Interactions"}
 Springer - Verlag Berlin - Heidelberg, 1998.
\bibitem{ake} D.S. Akerib {\it et al}. ( DAMA Collaboration ), Phys. Rev. Lett.{\bf
93}, 211301(2004).
%\bibitem{asa}  T. Asaka and M. Yamaguchi. Phys. Rev.{\bf D59} (1999) 125003.
%\bibitem{has} M. Hashimoto {\it et al}., Phys. Lett {\bf B447}, 44
%(1998).
%\bibitem{tl11} Daniel Arndt, Patrick J. Fox.
\end{thebibliography}
\end{document}